\documentclass{segabs}
\usepackage{amsmath}
\usepackage{mathtools}
\usepackage{amsfonts}
\usepackage{amssymb}
\usepackage{hyperref}

\begin{document}

%%%%%%%%%%%%%%%%%%%%%%%%%%%%%%%%%%%%%%%%%%%%%%%%%%%%%%%%%%%%%%%%%%%%%%%
% Cover Page begins here

\onecolumn % make sure you keep this coverpage as one column. In this template, we force the coverpage to be one column with this command and then switch to double column for the remaining of the paper with the \doublecolumn command. 

%\begin{description}[labelindent=-3cm,leftmargin=1cm,style=multiline]
\begin{itemize}
\item[\textbf{Citation}]{A. Mustafa, and G. AlRegib, “Joint Learning for Seismic Inversion: An Acoustic Impedance Estimation Case Study,” Expanded Abstracts of the SEG Annual Meeting , Houston, TX, Oct. 11-16, 2020.}

%\item[\textbf{DOI}]{\url{https://doi.org/10.1109/MSP.2017.2783449}}

\item[\textbf{Review}]{Date of presentation: 14 Oct 2020}

% \item[\textbf{Data and Code}]{\href{https://github.com/olivesgatech/Estimation-of-acoustic-impedance-from-seismic-data-using-temporal-convolutional-network}{[\underline{Github Link}]}}

% \item[\textbf{Bib}] {@incollection\{amustafa2019AI,\\
% title=Estimation of Acoustic Impedance from Seismic Data using Temporal Convolutional Network, \\
% author=Mustafa, Ahmad and AlRegib, Ghassan, \\
% booktitle=SEG Technical Program Expanded Abstracts 2019, \\
% year=2019, \\
% publisher=Society of Exploration Geophysicists\}
% } 

% Preprint sharing policy can vary depending on the publisher. Before posting a paper to arXiv, please specifically check the transaction/convference you are targeting. In some transactions, papers are usually added to arxiv after acceptance. Pubslishers usually allow the authors to share accepted version of their papers but not the final formatted version that is provided by the pubisher. In case of sharing preprints, publishers usually ask to add DOI and citation to the paper along with a copyright notice.

\item[\textbf{Copyrights}]{This is a preprint of a manuscript that has been accepted to be presented at the SEG Annual Meeting 2020. This preprint may strictly be used for research purposes only.}

\item[\textbf{Contact}]{\href{mailto:amustafa9@gatech.edu}{amustafa9@gatech.edu}  OR \href{mailto:alregib@gatech.edu}{alregib@gatech.edu}\\ \url{http://ghassanalregib.info/} \\ }
\end{itemize}

% Cover Page ends here

\thispagestyle{empty}
\newpage
\clearpage
\setcounter{page}{1}

\twocolumn
%%%%%%%%%%%%%%%%%%%%%%%%%%%%%%%%%%%%%%%%%%%%%%%%%%%%%%%%%%%%%%%%%%%%%%%%

\title{Joint Learning for Seismic Inversion: An Acoustic Impedance Estimation Case Study}

\renewcommand{\thefootnote}{\fnsymbol{footnote}} 

\author{Ahmad Mustafa\footnotemark[1]
  and Ghassan AlRegib, Center for Energy and Geo Processing (CeGP), School of Electrical and Computer Engineering, Georgia Institute of Technology}

\footer{Example}
\lefthead{Mustafa \& AlRegib}
\righthead{Joint Learning for Seismic Inversion}

\maketitle
\section{Summary}
Seismic inversion helps geophysicists build accurate reservoir models for exploration and production purposes. Deep learning-based seismic inversion works by training a neural network to learn a mapping from seismic data to rock properties using well log data as the labels. However, well logs are often very limited in number due to the high cost of drilling wells. Machine learning models can suffer overfitting and poor generalization if trained on limited data. In such cases, well log data from other surveys can provide much needed useful information for better generalization. We propose a learning scheme where we simultaneously train two network architectures, each on a different dataset. By placing a soft constraint on the weight similarity between the two networks, we make them learn from each other where useful for better generalization performance on their respective datasets. Using less than 3$\%$ of the available training data, we were able to achieve an average $r^{2}$ coefficient of 0.8399 on the acoustic impedance pseudologs of the SEAM dataset via joint learning with the Marmousi dataset.   

\section{Introduction}
Seismic inversion refers to the process of estimating rock properties in the subsurface. This allows geophysicists to build  accurate reservoir models for hydrocarbon exploration and production. While these properties can be measured directly at the well locations, they must be estimated using seismic data at the non-well locations. Classical seismic inversion usually works by starting with a smooth model of subsurface properties and forward modeling it to generate synthetic seismic data. The synthetic seismic is compared to the actual seismic and the difference between the two is used to update the model parameters. A detailed overview of classical seismic inversion methods is provided by \citep{Veeken2004SeismicIM}. \\

Deep learning, a subset of machine learning, has in the recent past led to ground breaking advancements in the field of Image classification  \citep{Krizhevsky2017}, object detection \citep{objdetect}, image segmentation \citep{segmentation}, image and video captioning \citep{captioning}, speech recognition \citep{speech}, and machine translation \citep{DBLP:conf/emnlp/ChoMGBBSB14}. The success of deep learning in computer vision and natural language processing domains has of late inspired geophysicists to replicate these successes in the field of seismic interpretation. Machine learning has been used to solve problems in salt body delineation \citep{haibinSaltbodyDetection, AsjadSaltDetection, AmirSaltDetection} , fault detection \citep{haibinFaultDetection, HaibinFaultDetection2}, facies classification \citep{YazeedFaciesClassification, YazeedFaciesWeakClassification}, and structural similarity based seismic image retrieval and segmentation \citep{YazeedStructurelabelPrediction}. 

Recently, there has been a lot of interest in developing deep learning-based workflows for seismic inversion. \citep{BiswasPhysicsGuidedCNN, DasCNNInversion} used Convolutional Neural Networks (CNNs) for estimating Acoustic and Elastic Impedance from seismic data. \cite{motazRNN1} and \cite{mustafaTCN} introduced sequence modelling-based neural networks based on Recurrent Neural Networks (RNNs) and Temporal Convolutional Network (TCN) respectively for estimation of various rock properties from seismic data. They demonstrated that such networks were more capable of learning temporal relationships between seismic traces for efficient rock property estimation. \citep{motazSemiSupervisedAcoustic, motazSemiSupervisedElastic} also showed how incorporating the forward model into the network architecture resulted in an implicit regularization of the network, thereby improving the quality of property estimations. 

All such methods are based upon learning a mapping from seismic data to well log measurements at the well positions, and then using the learned mapping to estimate the properties at the non-well positions. One limitation with such approaches is that they require a lot of labeled training data to achieve satisfactory generalization performance. However, most surveys have only a limited number of wells, due to the high cost of drilling them. This makes machine learning models prone to overfitting if trained on such limited well log data. One way of overcoming this is to use knowledge gained from learning on well logs from other surveys in estimating rock properties on the target survey. 

Transfer learning is a very popular machine learning framework that uses knowledge from a source dataset while training a machine learning model on the target dataset. It has been shown to help achieve better generalization performance and quicker convergence on the target dataset. It also results in less effort being expended to manually label training examples in the target survey, especially when it is costly and time consuming. For a comprehensive review of transfer learning methodologies that have been used in the past, refer to \citep{transferlearning}.     

In this paper, we propose a transfer learning scheme for seismic inversion that jointly learns on multiple datasets using 
identical copies of the same network architecture. In addition to optimizing the losses on their respective datasets, we also impose a soft constraint on the weights of the network copies to be similar to each other. This effectively results in a knowledge sharing scheme where the two networks are learning from each other where it is mutually beneficial while being able at the same time to adapt to the specific nature of their respective dataset.

\section{Methodology}
\subsection{2-D Temporal Convolutional Network}
\begin{figure*}[htbp]
\centering
\includegraphics[width=2\columnwidth]{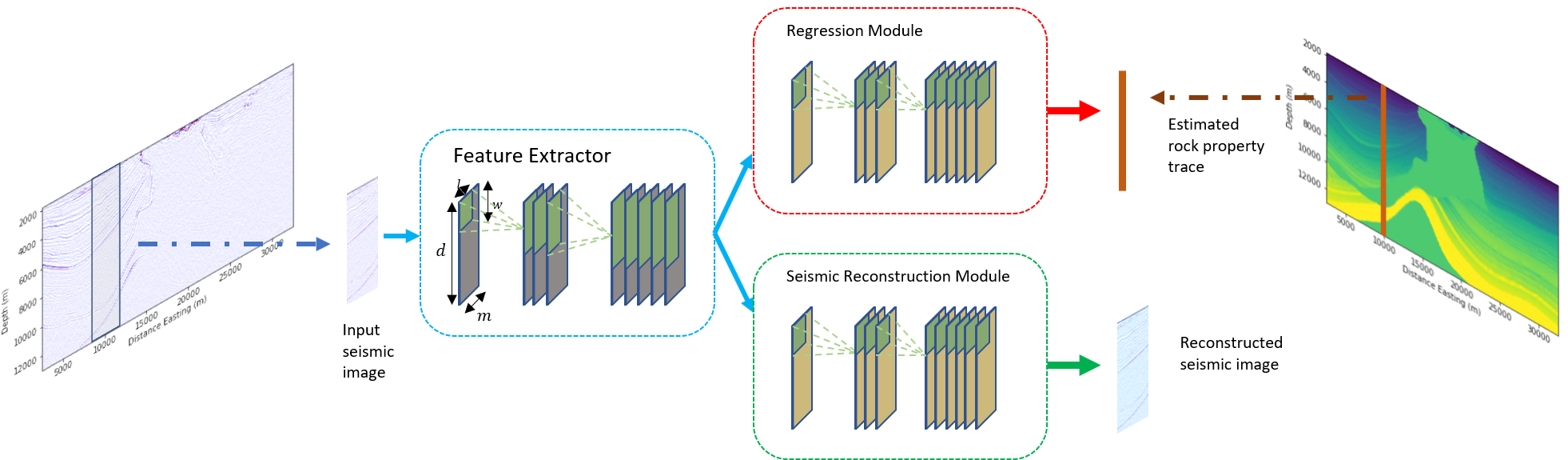}
\caption{The Architecture uses a series of 2-D Temporal Convolutional Blocks to extract features from the input. The input is a 2-D patch of seismic data centered at the well position. The output of the Feature Extractor is fed simultaneously into the regression module and the reconstruction module for the estimation of rock property and reconstruction of seismic input respectively.}
\label{fig:architecture}
\end{figure*}

As mentioned beforehand, our algorithm employs the use of two dimensional Temporal Convolutional Blocks for estimating rock properties from seismic data. The architecture is shown in Figure ~\ref{fig:architecture}. The Architecture consists of a feature extractor module that uses a series of 2-D convolutional kernels to extract increasingly abstract features from the input seismic image. The convolutional kernels in this module use an exponentially increasing dilation factor in depth while staying constant in width. Using increasingly dilated convolutions in depth allows us to model input seismic data temporally for efficiently capturing long term dependencies, leading to better estimation of the desired rock property. The kernel being 2-D allows us to inject spatial context awareness into our network estimations. The output of the feature extractor block is fed simultaneously into a Regression module and a Reconstruction module. The latter is responsible for reconstructing the input seismic image while the former outputs the desired rock property. This is an example of multi-task learning via representation sharing, where multiple tasks (output estimation and input reconstruction in this case) are learnt simultaneously in the hope that the network can learn more generalizable feature representations, leading to better performance on all tasks. This is especially the case when the tasks are highly related to each other.

\subsection{Soft Weight Sharing}
\begin{figure}[htbp]
\centering
\includegraphics[width=\columnwidth]{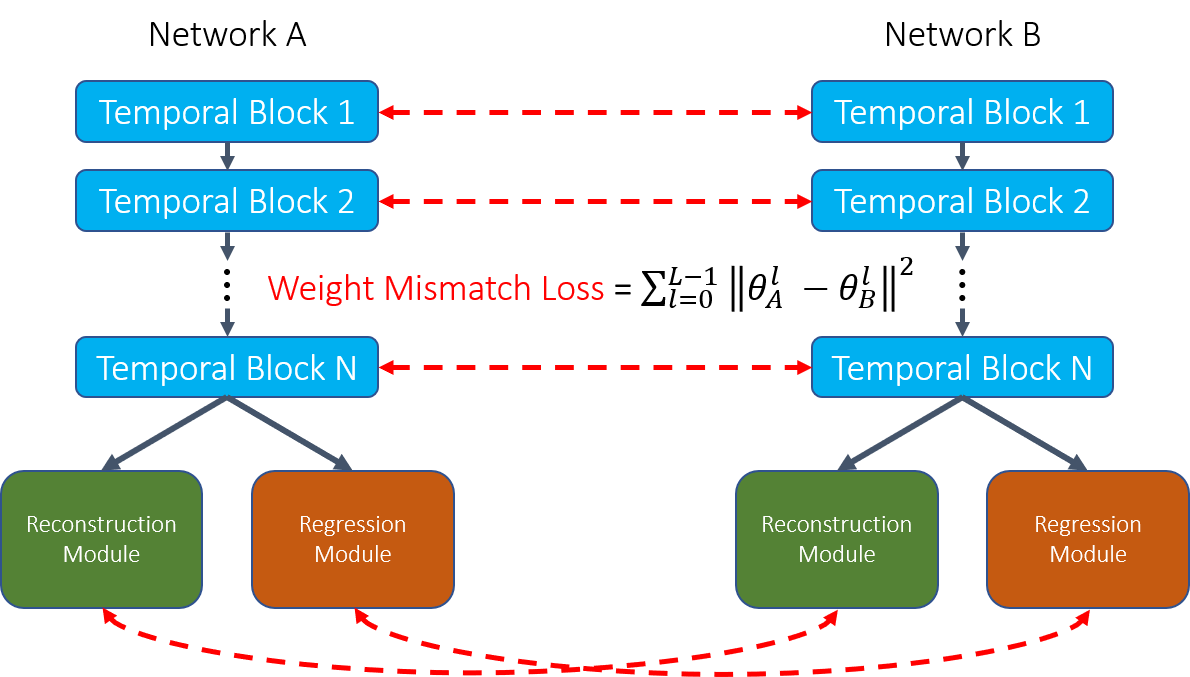}
\caption{The two architectures are constrained to have weights in all corresponding layers close to each other in the L2 norm sense.}
\label{fig:weight_sharing}
\end{figure}

As discussed before, another major component of our deep learning based-seismic inversion algorithm is learning from related datasets for rock property prediction. This is achieved by simultaneously training identical copies of the same architecture, one for each dataset. Each network receives a batch of input training examples from its respective dataset, processes them to get the outputs, and uses the corresponding ground-truths to backpropagate the losses through the network to update network weights. In addition to this, we also force the network weights in all corresponding layers to be close to each other in the L2 norm sense. By doing this, we effectively bias the networks to search the parameter space for a solution where the architecture will generalize better to inputs sampled from different distributions. However, by not constraining the weights to be exactly the same, each copy of the architecture is also free to find the optimal set of weights for its dataset in the vicinity of this solution space. Moreover, in the situation where the two datasets are very different from each other and learning on one will not help the other, the networks can choose to not learn from each other at all. The process is illustrated in Figure ~\ref{fig:weight_sharing}. Consider the two networks to be represented by $\mathcal{F}$ and $\mathcal{G}$ respectively. Both $\mathcal{F}$ and $\mathcal{G}$ consist of trainable weights organized into a set of $L$ convolutional layers. Consider $\theta_{A}^{l}$ to be the weight tensor in the $l$-th layer in network $A$, where $l\in [0, L-1]$. Then both $\mathcal{F}$ and $\mathcal{G}$ can be represented as follows:
\begin{equation}
    \mathcal{F} = [\theta_{F}^{0}, \theta_{F}^{1},\cdots, \theta_{F}^{L-1}]
    \label{eq:network1}
\end{equation}

\begin{equation}
    \mathcal{G} = [\theta_{G}^{0}, \theta_{G}^{1},\cdots, \theta_{G}^{L-1}] 
    \label{eq:network2}
\end{equation}

The Weight Mismatch Loss is then defined as:
\begin{equation}
    l_{WML} = \sum_{l=0}^{L-1} \|\theta_{\mathcal{F}}^{l} - \theta_{\mathcal{G}}^{l}\|_{2}^{2}
    \label{eq:weight_mismatch}
\end{equation}
 
% If left like this, each network would learn independently of the other to optimize on its respective dataset. In the case where we have very few training examples, the networks could easily overfit on the training examples.  

\subsection{Network Training}
Consider $D_{1} = \{X_{1}, Y_{1}\}$ and $D_{2} = \{X_{2}, Y_{2}\}$ to represent two datasets, where the subscript refers to the dataset. $X = \{x^{1}, ..., x^{N}| x^{i} \in \mathbb{R}^{d \times m}\}$ represents the collection of $N$ seismic images in a dataset, where each $x^{i}$ is a $d\times m$ dimensional image. $d$ refers to the depth of the image while $m$ is the width. $Y = \{y^{1}, ..., y^{N}|y^{i}\in\mathbb{R}^{d}\}$ refers to collection of well log properties corresponding to each $x^{i} \in X$, where each $y^{i}$ is a $d$ dimensional rock property trace.
A batch of seismic images from each dataset is processed by its respective network to get the estimated well properties, $\hat{y}^{i}$ as well as the reconstructed seismic images, $\hat{x}^{i}$ as shown below:

\begin{equation}
    \hat{y}_{1}^{i}, \hat{x}_{1}^{i} = \mathcal{F}_{\Theta}(x_{1}^{i}) 
\label{eq:4}
\end{equation}

\begin{equation}
    \hat{y}_{2}^{i}, \hat{x}_{2}^{i} = \mathcal{G}_{\Theta}(x_{2}^{i}) 
\label{eq:5}
\end{equation}

The regression and reconstruction losses are then defined as:
\begin{equation}
    l_{reg} = \frac{1}{N_{1}}\sum_{i=1}^{N_{1}}\|\hat{y_{1}}^{i} - y_{1}^{i}\|_{2}^{2} + \frac{1}{N_{2}}\sum_{i=1}^{N_{2}}\|\hat{y}_{2}^{i} - y_{2}^{i}\|_{2}^{2}
\end{equation}
and 
\begin{equation}
    l_{recon} = \frac{1}{N_{1}}\sum_{i=1}^{N_{1}}\|\hat{x_{1}}^{i} - x_{1}^{i}\|_{2}^{2} + \frac{1}{N_{2}}\sum_{i=1}^{N_{2}}\|\hat{x}_{2}^{i} - x_{2}^{i}\|_{2}^{2}
\end{equation}

where $N_{1}$ and $N_{2}$ are the batch sizes in the two datasets.
The total loss is then obtained as:
\begin{equation}
    \textrm{Total Loss} = l_{reg} + l_{recon} + \alpha\times l_{WML},
\end{equation}
where $\alpha$ is a hyperparameter that controls the influence of the weight mismatch loss on the training of the two networks. Over each training iteration, the loss obtained above is backpropagated through both networks and the weights updated to reduce the training error at the next iteration. If $\alpha$ is set too high, it forces the networks to look for the same solution, that might not be optimized for each dataset individually. If $\alpha$ is set too low, it makes the training of the two networks effectively independent of each other. An intermediate value for $\alpha$ results in the two networks learning from each other when it is useful for optimization on their own datasets, and ignoring each other when knowledge sharing is not useful. 

\section{Results and Discussion}
\begin{figure}[htbp]
\centering
\includegraphics[width=\columnwidth]{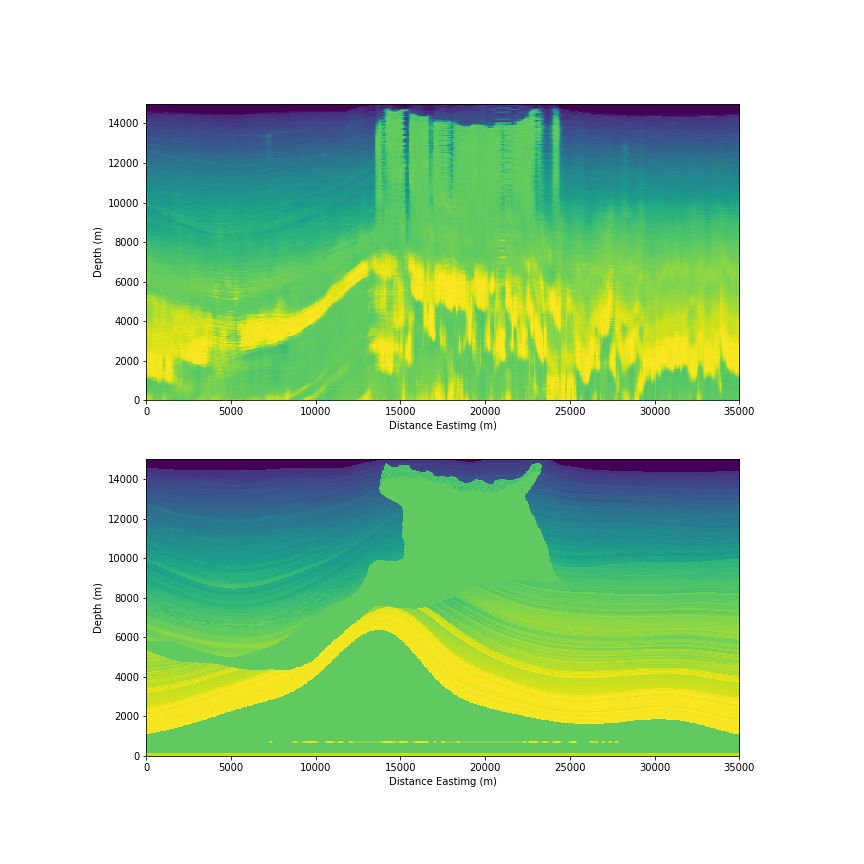}
\caption{Estimated Acoustic Impedance section (top) vs the groundtruth (bottom).}
\label{fig:seam}
\end{figure}

We demonstrate our workflow for the estimation of Acoustic Impedance from poststack, migrated seismic data on the open source Marmousi and SEAM datasets. We set up a network Architecture as shown in Figure ~\ref{fig:architecture} for each dataset, and train them jointly for 900 epochs. We use ADAM \citep{kingma2014adam} as the optimization algorithm, which adaptively sets the learning rate during the progression of training. We also impose a weight decay constraint of 0.001 which helps preventt overfitting by constraining the L2 norm of the network weights. We uniformly sample 12 acoustic impedance pseudologs and their corresponding seismic trace data from the crossline section in SEAM located at North 23900m. For Marmousi, we sample uniformly 51 acoustic impedance pseudologs and the corresponding seismic data in the dataset. Marmousi is a synthetic dataset with the seismic data generated by simple convolutional forward modeling, while SEAM     
contains migrated seismic data made to simulate real world acquisition conditions and artifacts. This makes it a much harder dataset to learn on, especially with only a limited number of pseudologs available. We use a greater number of pseudologs in Marmousi to provide the network training on SEAM with sufficient information to learn from. The results of this training scheme are illustrated in Figure ~\ref{fig:seam}. One can clearly see that our algorithm is able to delineate with sufficient detail, the vertical variations in acoustic impedance, especially in the left half of the section. The top of the salt dome has also been marked out to a high degree of accuracy, despite us not having access to many pseudologs there. We have also been able to mark out the top and the bottom of the high impedance arch occurring around a depth of 12000m to a reasonable degree of detail. One can see that estimation get noisier in the bottom-right portion of the section. This is to be expected since the seismic data in these regions is extremely weak and sometimes not receptive at all to variations in acoustic impedance. Despite this, our algorithm is still able to capture the general increasing trend in acoustic impedance values well. Figure 4 shows some individual acoustic impedance traces extracted from both the estimated and ground truth acoustic impedance sections at select positions and overlaid on top of each other. As explained before, the two largely agree with each other for the most part, except around a depth of 12000m where we have a sudden jump in the value for acoustic impedance, along with a weakened seismic data in the seismic section. The $r^{2}$ coefficient, also called the coefficient of determination, gives information about the goodness of fit of a model. Given a set of observed values and a corresponding set of predicted values, the $r^{2}$ coefficient is an indicator of how well the predicted values approximate the real ones. A value of 1 indicates a perfect fit. We calculated the average $r^{2}$ coefficient between the estimated and ground truth acoustic impedance sections on SEAM, and it turns out to be 0.8399, which indicates that our model has been able to model acoustic impedance well, given that we only had 12 training samples to train it with, which is around 2$\%$ of the total available training data.    

\begin{figure}[htbp]
\centering
\includegraphics[width=\columnwidth]{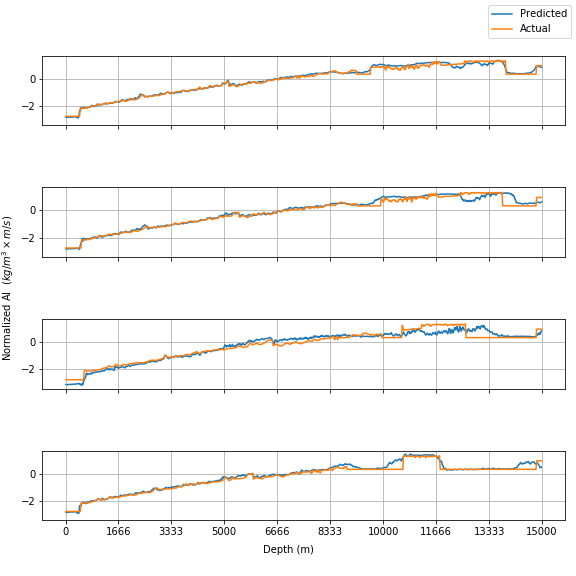}
\caption{Trace plots for select positions in both the estimated and ground-truth acoustic impedance sections.}
\label{fig:traces}
\end{figure} 

\section{Conclusion}
In this work, we demonstrate a deep learning-based seismic inversion workflow where we jointly train identical copies of a neural network on two different datasets. We show how, by placing a soft constrain on the network weights to be similar, we allow the transfer of useful knowledge to take place from one dataset to the other while simultaneously letting the networks adapt to their specific datasets. An important implication of this workflow is that one need not use a large number of training examples in either, since mutual information between the them would serve to compensate for that. Another implication for this work is that one can scale this workflow to any number of datasets. We demonstrate the utility of this approach through estimating Acoustic Impedance on the SEAM dataset, although the methodology would be equally valid for other rock properties.

\bibliographystyle{seg}  % style file is seg.bst
\bibliography{example}

\newcommand{\SortNoop}[1]{}
\begin{thebibliography}{}
\itemsep0pt

\bibitem[Alaudah et~al., 2019a]{YazeedStructurelabelPrediction}
Alaudah, Y., M. Alfarraj, and G. AlRegib,  2019a, Structure label prediction
  using similarity-based retrieval and weakly supervised label mapping:
  GEOPHYSICS, {\bf 84}, V67--V79.

\bibitem[Alaudah et~al., 2019b]{YazeedFaciesClassification}
Alaudah, Y., P. Michałowicz, M. Alfarraj, and G. AlRegib,  2019b, A
  machine-learning benchmark for facies classification: Interpretation, {\bf
  7}, SE175--SE187.

\bibitem[Alaudah et~al., 2019c]{YazeedFaciesWeakClassification}
Alaudah, Y., M. Soliman, and G. AlRegib,  2019c, Facies classification with
  weak and strong supervision: A comparative study: Presented at the {SEG}
  Technical Program Expanded Abstracts 2019, Society of Exploration
  Geophysicists.

\bibitem[Alfarraj and AlRegib, 2018]{motazRNN1}
Alfarraj, M., and G. AlRegib,  2018, Petrophysical-property estimation from
  seismic data using recurrent neural networks: Presented at the {SEG}
  Technical Program Expanded Abstracts 2018, Society of Exploration
  Geophysicists.

\bibitem[Alfarraj and AlRegib, 2019a]{motazSemiSupervisedAcoustic}
--------, 2019a, Semi-supervised learning for acoustic impedance inversion:
  Presented at the {SEG} Technical Program Expanded Abstracts 2019, Society of
  Exploration Geophysicists.

\bibitem[Alfarraj and AlRegib, 2019b]{motazSemiSupervisedElastic}
--------, 2019b, Semisupervised sequence modeling for elastic impedance
  inversion: Interpretation, {\bf 7}, SE237--SE249.

\bibitem[Amin et~al., 2017]{AsjadSaltDetection}
Amin, A., M. Deriche, M.~A. Shafiq, Z. Wang, and G. AlRegib,  2017, Automated
  salt-dome detection using an attribute ranking framework with a
  dictionary-based classifier: Interpretation, {\bf 5}, SJ61--SJ79.

\bibitem[Biswas et~al., 2019]{BiswasPhysicsGuidedCNN}
Biswas, R., M.~K. Sen, V. Das, and T. Mukerji,  2019, Prestack and poststack
  inversion using a physics-guided convolutional neural network:
  Interpretation, {\bf 7}, SE161--SE174.

\bibitem[{Chen} et~al., 2018]{segmentation}
{Chen}, L., G. {Papandreou}, I. {Kokkinos}, K. {Murphy}, and A.~L. {Yuille},
  2018, Deeplab: Semantic image segmentation with deep convolutional nets,
  atrous convolution, and fully connected crfs: IEEE Transactions on Pattern
  Analysis and Machine Intelligence, {\bf 40}, 834--848.

\bibitem[Cho et~al., 2014]{DBLP:conf/emnlp/ChoMGBBSB14}
Cho, K., B. van Merrienboer, {\c{C}}. G{\"{u}}l{\c{c}}ehre, D. Bahdanau, F.
  Bougares, H. Schwenk, and Y. Bengio,  2014, Learning phrase representations
  using {RNN} encoder-decoder for statistical machine translation: Proceedings
  of the 2014 Conference on Empirical Methods in Natural Language Processing,
  {EMNLP} 2014, October 25-29, 2014, Doha, Qatar, {A} meeting of SIGDAT, a
  Special Interest Group of the {ACL}, {ACL}, 1724--1734.

\bibitem[Das et~al., 2019]{DasCNNInversion}
Das, V., A. Pollack, U. Wollner, and T. Mukerji,  2019, Convolutional neural
  network for seismic impedance inversion: {GEOPHYSICS}, {\bf 84}, R869--R880.

\bibitem[Di and AlRegib, 2019]{haibinFaultDetection}
Di, H., and G. AlRegib,  2019, Semi-automatic fault/fracture interpretation
  based on seismic geometry analysis: Geophysical Prospecting, {\bf 67},
  1379--1391.

\bibitem[Di et~al., 2018]{haibinSaltbodyDetection}
Di, H., M. Shafiq, and G. AlRegib,  2018, {Multi-attribute k-means clustering
  for salt-boundary delineation from three-dimensional seismic data}:
  Geophysical Journal International, {\bf 215}, 1999--2007.

\bibitem[Di et~al., 2019]{HaibinFaultDetection2}
Di, H., M.~A. Shafiq, Z. Wang, and G. AlRegib,  2019, Improving seismic fault
  detection by super-attribute-based classification: Interpretation, {\bf 7},
  SE251--SE267.

\bibitem[{Graves} et~al., 2013]{speech}
{Graves}, A., A. {Mohamed}, and G. {Hinton},  2013, Speech recognition with
  deep recurrent neural networks: 2013 IEEE International Conference on
  Acoustics, Speech and Signal Processing, 6645--6649.

\bibitem[Kingma and Ba, 2014]{kingma2014adam}
Kingma, D.~P., and J. Ba,  2014, Adam: A method for stochastic optimization.

\bibitem[Krizhevsky et~al., 2017]{Krizhevsky2017}
Krizhevsky, A., I. Sutskever, and G.~E. Hinton,  2017, {ImageNet}
  classification with deep convolutional neural networks: Communications of the
  {ACM}, {\bf 60}, 84--90.

\bibitem[Mustafa et~al., 2019]{mustafaTCN}
Mustafa, A., M. Alfarraj, and G. AlRegib,  2019, Estimation of acoustic
  impedance from seismic data using temporal convolutional network: Presented
  at the {SEG} Technical Program Expanded Abstracts 2019, Society of
  Exploration Geophysicists.

\bibitem[{Pan} and {Yang}, 2010]{transferlearning}
{Pan}, S.~J., and Q. {Yang},  2010, A survey on transfer learning: IEEE
  Transactions on Knowledge and Data Engineering, {\bf 22}, 1345--1359.

\bibitem[Ren et~al., 2015]{objdetect}
Ren, S., K. He, R. Girshick, and J. Sun,  2015, Faster r-cnn: Towards real-time
  object detection with region proposal networks: Proceedings of the 28th
  International Conference on Neural Information Processing Systems - Volume 1,
  MIT Press, 91–99.

\bibitem[Shafiq et~al., 2017]{AmirSaltDetection}
Shafiq, M.~A., Z. Wang, G. AlRegib, A. Amin, and M. Deriche,  2017, A
  texture-based interpretation workflow with application to delineating salt
  domes: Interpretation, {\bf 5}, SJ1--SJ19.

\bibitem[Veeken and Silva, 2004]{Veeken2004SeismicIM}
Veeken, P., and M.~D. Silva,  2004, Seismic inversion methods and some of their
  constraints: First Break, {\bf 22}.

\bibitem[{Vinyals} et~al., 2015]{captioning}
{Vinyals}, O., A. {Toshev}, S. {Bengio}, and D. {Erhan},  2015, Show and tell:
  A neural image caption generator: 2015 IEEE Conference on Computer Vision and
  Pattern Recognition (CVPR), 3156--3164.

\end{thebibliography}

\end{document}